\documentclass[lettersize,journal]{IEEEtran}
\usepackage[utf8]{inputenc}
\usepackage[cmex10]{amsmath}
\usepackage{upgreek}
\usepackage{booktabs}
\usepackage{epsfig}
\usepackage{latexsym}
\usepackage{multirow}
\usepackage{stfloats}
\usepackage{epstopdf}
\usepackage{color}  
\usepackage{tabularx} 
\usepackage{algorithm}
\usepackage{algpseudocode}
\usepackage{amssymb}
\usepackage{enumerate}
\usepackage{array}
\graphicspath{{./Figures/}}
\usepackage{color}
\usepackage{bbm}
\usepackage{bm}
\usepackage{cite}
\usepackage[tight,footnotesize]{subfigure}
\usepackage{balance}
\usepackage{mathrsfs}
\usepackage{verbatim}
\usepackage{dsfont}
\usepackage{verbatim}
\usepackage{setspace}
\usepackage{diagbox}
\usepackage{multicol}
\usepackage{environ}
\usepackage{tikz}
\usepackage{amsmath}
\usepackage{stfloats}
\usepackage{algorithm}
\usepackage{algpseudocode}
\usepackage{amsmath}
\usepackage{graphics}
\usepackage{epsfig}
\usepackage{caption}
\allowdisplaybreaks[4]
\usepackage{amsmath}
\usepackage{amsthm}
\usepackage{authblk}

\def\BibTeX{{\rm B\kern-.05em{\sc i\kern-.025em b}\kern-.08em
		T\kern-.1667em\lower.7ex\hbox{E}\kern-.125emX}}
\begin{document}
\title{Terahertz Aerospace Communications: Enabling Technologies and Future Directions}
	
\author{Weijun Gao, Chong Han, Zhi Chen, Yong Chen, Yuanzhi He, and Wenjun Zhang
\thanks{
Weijun~Gao is with the Terahertz Wireless Communications (TWC) Laboratory, Shanghai Jiao Tong University, Shanghai 200240, China (email:~gaoweijun@sjtu.edu.cn). 

Chong~Han is with the Terahertz Wireless Communications (TWC) Laboratory, Department of Electronic Engineering and Cooperative Medianet Innovation Center (CMIC), Shanghai Jiao Tong University, Shanghai 200240, China (email:~chong.han@sjtu.edu.cn). 

Zhi~Chen is with the National Key Laboratory of Science and Technology on Communications (NCL), University of Electronic Science and Technology of China, Chengdu 611731, China (email: chenzhi@uestc.edu.cn).

Yong~Chen is with Chengdu Fluid Dynamics Innovation Center, Chengdu, China (email:~literature\_chen@nudt.edu.cn).

Yuanzhi~He is with Institute of Systems Engineering, Academy of Military Sciences, Beijing 100141, China (email: he\_yuanzhi@126.com).

Wenjun~Zhang is with Department of Electronic Engineering and Cooperative Medianet Innovation Center (CMIC), Shanghai Jiao Tong University, Shanghai 200240, China (email:~zhangwenjun@sjtu.edu.cn). 
}
}

	\maketitle
	\thispagestyle{empty}
	\begin{abstract}
To achieve ubiquitous connectivity in next-generation networks through aerospace communications while maintaining high data rates, Terahertz (THz) band communications (0.1-10~THz)  with large continuous bandwidths are considered a promising candidate technology. However, key enabling techniques and practical implementations of THz communications for aerospace applications remain limited. In this paper, the wireless channel characteristics, enabling communication techniques, and networking strategies for THz aerospace communications are investigated, aiming to assess their feasibility and encourage future research efforts toward system realization. 
Specifically, the wireless channel characteristics across various altitudes and scenarios are first analyzed, focusing on modeling the interaction between the THz wave and the external environment, from ground to outer space. Next, key enabling communication technologies, including multiple-input multiple-output (MIMO) technique, beam alignment and tracking, integrated communication and radar sensing (ICARS), and resource allocation for networking are discussed. Finally, the existing challenges and possible future directions are summarized and discussed.  
	\end{abstract}
	
    \section{Introduction}
    Despite the evolving transmission capabilities of wireless communication networks in urban areas, the ever-increasing demand for ubiquitous connectivity to deliver wireless services anytime and anywhere drives the need to extend connections to remote areas, such as rural areas, mountains, deserts, and oceans. One promising solution is radio access from space, which has minimal blockage and offers extensive coverage. This has sparked significant interest in the research of aerial and space communications. 
    As aerospace device technologies evolve, the number of unmanned aerial vehicles in low-altitude regions, as well as satellites in low Earth orbits (LEO), medium Earth orbit (MEO), and geosynchronous orbit (GEO), has increased dramatically. 
    In the past decade, numerous UAV and satellite projects have been undertaken by commercial organizations such as SpaceX in America, Telesat in Canada, and OneWeb in the United Kingdom. Additionally, the 3rd Generation Partnership Project (3GPP) has launched research programs on non-terrestrial networks (NTN) to develop enabling communication techniques for aerospace communications.
    These UAVs and satellites serve as communication base stations or relays, making them key components of future communication networks~\cite{elbir2024terahertz,gao2024attenuation}.

    Despite the increased coverage supported by UAVs and satellites, the limited bandwidth resources in the conventional frequency band bound the maximum data rate. According to the request for high-speed data connection anywhere and anytime, the Terahertz (THz) band, with frequency ranging from $0.1~\textrm{THz}$ to $10~\textrm{THz}$ and wavelength from $3~\textrm{millimeters}$ to $30~\textrm{micrometers}$, has been envisioned as a key technology to boost the achievable rate in future aerospace communications~\cite{mao2022terahertz,alqaraghuli2023road}.
    The advantages of THz communications for the aerospace scenarios are summarized as follows.
    
    First, compared with low-frequency band communications such as microwave or millimeter-wave band, the large continuous bandwidth of the THz band on the order of multi-hundred gigahertz can enable 10-100 times faster wireless data transfer between resource-constrained airplanes and satellites. 
    Second, compared with optical communications where even higher data rates than the THz band are achievable, THz communications have better resilience to sunlight than free-space optical (FSO) communications, which address the challenge of link outages when being interfered with the sunlight for aerospace communications. 
    Moreover, THz band communications possess better flexibility due to the use of an ultra-massive multiple-input multiple-output (UM-MIMO) antenna array, which can steer the beam direction electronically by phase shifters rather than physically altering the antenna direction. This is more suitable for aerospace communications since the constellation of satellites and the movement of UAVs and airplanes are changing frequently and the physical movement of the antenna consumes additional power.
    
    Besides the advantages of the THz band over other frequency bands, the aerospace communication scenarios are more suitable for the THz band than terrestrial ones. One significant advantage is the reduced molecular absorption effect and other attenuation effects. For example, the water vapor density is around a typical value of $7.5\textrm{g}/\textrm{m}^3$ in terrestrial environment~\cite{ITU-RP.676}, which severely limits the maximum transmission distance at certain frequencies such as $0.183~\textrm{THz}$ and $0.557~\textrm{THz}$. Such limitation is alleviated and even disappears in high-altitude regions and space since the atmosphere becomes thinner at higher altitudes.

    Despite these aforementioned advantages, THz aerospace communications still face great challenges in the THz link establishment and realization of future applications. 
    In this paper, we elaborate on the challenges, existing research progress, and remaining open problems for THz aerospace communications. 
    Specifically, the wireless channel characteristics for THz aerospace are first investigated, which theoretically verifies the feasibility of THz aerospace communication.
    Second, some key enabling techniques for the THz band are described. 
    To be concrete, the
    ultra-massive multiple-input multiple-output (UM-MIMO) and intelligent surface techniques, which are advocated to compensate for the large path loss over kilometers of propagation distances in aerospace communications, are investigated. 
    Then, beam alignment and tracking techniques are explored to maintain the high antenna gain in this misalignment-sensitive scenario. Moreover, integrated communication and radar sensing~(ICARS) and the advanced resource allocation techniques for THz aerospace networking are investigated to jointly improve the performance of THz universe sensing and data back-haul. 

    The remainder of the paper is organized as follows.
    The THz wireless channel models for low-altitude atmosphere, high-altitude atmosphere, and outer space are discussed and analyzed in Sec~\ref{sec:channel_modeling}. Key enabling technologies for THz aerospace communications including MIMO and intelligent surfaces, beam alignment and tracking, ICARS, and resource allocation are elaborated in Sec.~\ref{sec:enabling}. Furthermore, Sec.~\ref{sec:future} discusses the existing open problems and possible research directions for THz aerospace communications. Finally, this paper is concluded in Sec.~\ref{sec:concl}
    
    \begin{figure*}
        \centering
        \includegraphics[width=0.8\linewidth]{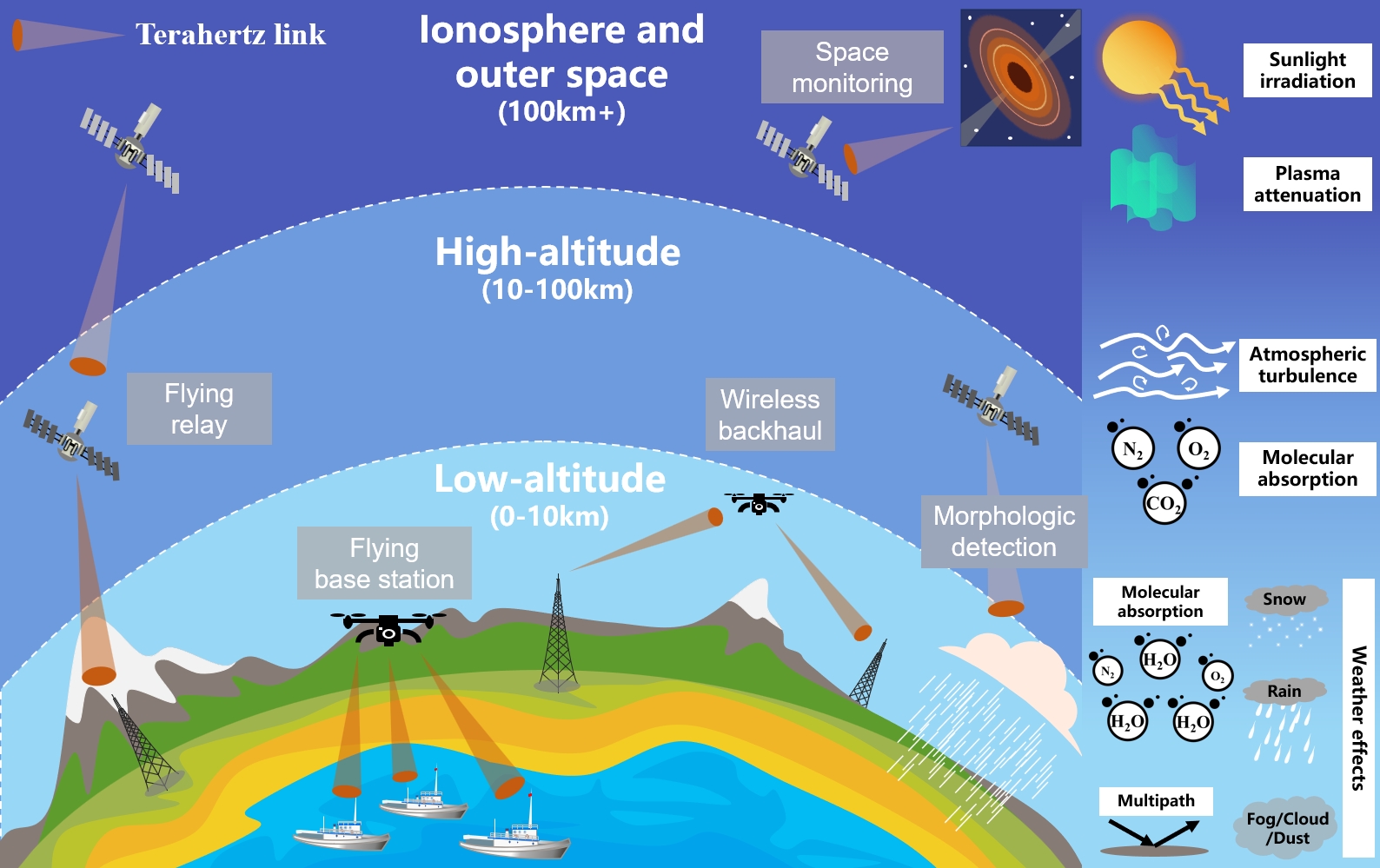}
        \caption{Scenario model for THz aerospace communications.}
        \label{fig:system-model}
    \end{figure*}
	
	\section{Channel Modeling for Terahertz Aerospace Communications}\label{sec:channel_modeling}

Unlike THz terrestrial wireless channels, where multi-path propagation effects such as reflection, diffraction, and scattering dominate, THz aerospace channels are primarily characterized by the interaction of line-of-sight (LoS) THz rays with the external environment due to the absence of spatial obstacles like walls or buildings~\cite{wang20226g}. The propagation environment significantly varies across different altitude regions, as illustrated in Fig.~\ref{fig:system-model}. In the troposphere ($0-10~\textrm{km}$), the high density of water vapor and severe weather effects, including rain and clouds, play a critical role in determining signal attenuation and distortion. In the stratosphere and mesosphere ($10-100~\textrm{km}$), the absence of aerosols shifts the focus to medium inhomogeneity caused by atmospheric turbulence, which disrupts THz wave propagation. Finally, in the ionosphere and outer space (above $100~\textrm{km}$), plasma effects dominate, introducing significant attenuation and dispersion. At these altitudes, the Doppler effect becomes critical, particularly for high-velocity satellite communications and inter-satellite links.
    
    \subsection{Low-altitude Aerial Channel Modeling}
    In low-altitude aerial communications up to about $10~\textrm{km}$, the multi-path effect near the ground, the molecular absorption effect due to water vapor, and attenuation due to weather effects need to be accounted for. 
        First, although there are few or even no multi-paths in high-altitude regions, THz channels in near-ground regions may possess some multi-paths. In~\cite{li2021ray}, simulation results demonstrate that the K-factor capturing the power ratio of the LoS path and other multi-path components are 0.39, 11.65, 49.10, for an altitude at $50~\textrm{m}$, $100~\textrm{m}$, $150~\textrm{m}$, respectively, which shows a decreasing trend with a rising altitude. Therefore, we just need to consider multi-path communications only when the communication altitude is in near-ground regions. One take-away lesson due to the lack of multi-paths is that as the altitude increases, the maintenance of the LoS path against blockage and the necessity of precise beam alignment and tracking becomes more significant. This is because if the LoS path is blocked or not perfectly aligned with each other, there are no alternative routes to maintain signal transmission.
    
    Second, the molecular absorption effect of water vapor is significant in low-altitude regions.
    When THz signals propagate through the channel medium, small polarized molecules (mainly water vapor in low-altitude regions) can be excited by the THz EM wave and therefore lead to additional absorption loss for THz wave propagation. This phenomenon is referred to as the molecular absorption effect. The molecular absorption effect is one of the most significant and special channel propagation features of THz communications, due to its triple selectivity~\cite{han2022molecular}, including distance dependency, frequency selectivity, and environment impact. Unlike free-space path loss which is linearly dependent on the square of distance and frequency, the molecular absorption loss is exponentially dependent on distance and shows a complicated dependency with frequency. Moreover, as the water vapor density decreases exponentially with altitude, the molecular absorption loss decreases dramatically as altitude increases.
    For THz aerospace communications, the triple selectivity implies that only THz communications in low-altitude regions need to consider the influence of molecular absorption, where the entire THz spectral band is separated into many discontinuous frequency windows. Only the so-called spectral windows can be used to achieve an acceptably high data rate. However, as the altitude increases, the range of spectral windows becomes wider and gradually meets with each other. 
    
    Third, the attenuation due to weather effects necessitates consideration.
    Weather conditions exist in the troposphere below the altitude of about $10~\textrm{km}$.
    Existing attenuation models for different weather effects are empirical and based on real measurement datasets. Recently, a collision-based model has been proposed as a universal model unifying these attenuation effects~\cite{yang2024universal}. According to the wave-particle duality, the THz wave propagation can be viewed as a stream of THz photons traveling through the medium, and the attenuation is due to the collision between these photons and medium particles including raindrops, fog droplets, sands, and snowflakes.     
    It is discovered that the two dominant factors affecting the attenuation of weather conditions are the size and material of the weather aerosols.
    
    On one hand, according to the thermodynamics and meteorology, the meteorological parameters such as temperature, pressure, rainfall rate, visibility, and snowfall rates are acquired by sensors and monitors. Then, by applying the thermodynamic theory the number distribution of aerosols is discovered. On the other hand, the intensity of absorption or scattering as the result of collision is modeled based on Rayleigh or Mie scattering theory.
    As a take-away lesson, the rain attenuation effect is more significant than other weather effects. This is due to the millimeter-level particle diameter of rain droplets that appear on the same order of magnitude with THz wavelengths. 
    Moreover, all these weather effects show a cut-off relationship with altitude. For example, the rainfall rate on rainy days at altitudes under the cumulonimbus is approximately uniform, while the rainfall rate above the cumulonimbus is zero. This is why the altitude of cumulonimbus at $10~\textrm{km}$ is taken as the boundary of low- and high-altitude regions. 

    \subsection{High-Altitude Aerial Channel Modeling}
    In high-altitude regions in the stratosphere and higher, no weather aerosols are contributing to additional attenuation. Based on the dataset in the high-resolution transmission molecular absorption database (HITRAN), the molecular absorption loss led by oxygen is the second-dominant one, with about six orders of magnitude less than that of water vapor. Moreover, the frequency selectivity of the molecular absorption by oxygen is less significant than that by water vapor but still stronger than Mie or Rayleigh scattering for weather effects. Moreover, due to the small magnitude of the molecular absorption of oxygen, this effect only needs to be considered for extremely large-distance communications.
    
    Besides molecules or aerosols in the atmosphere, the temperature and pressure inhomogeneity of air led by atmospheric turbulence is another key factor for THz wave propagation. 
    Due to the irradiation of sunlight, the temperatures of the Earth's surface and the atmosphere are different. This creates chaotic airflow in the atmosphere, referred to as atmospheric turbulence. 
    The random temperature and pressure fluctuation of the air brings random fluctuation of the refractive index of air and thereby leads to turbulent fading (known as scintillation in FSO communications) and attenuation. This effect is weak for microwave and millimeter-wave frequency bands but significant for free-space optical communication systems, since it is discovered that the scintillation loss is proportional to $f^{7/6}$, where $f$ denotes the carrier frequency. 
    In the THz band, which is located between the millimeter-wave band and free-space optical band, whether atmospheric turbulence can lead to non-negligible propagation loss remains an unanswered question.
    
    Due to the lack of experimental data regarding the influence of atmospheric turbulence on THz wave propagation, one of the alternative approaches is to theoretically model the turbulence-induced attenuation by extending the conclusions from the FSO band to the THz band, since the FSO channel model in turbulent media has been well studied.
    The channel modeling of the THz turbulent channel has two phases. 
    To start with, we need to model the atmosphere turbulence, especially the distribution of the refractive index of air in turbulence. Based on Kolmogorov's theory, the statistic of the inhomogeneity of the turbulent propagation medium is characterized by the refractive index structure constant (RISC), defined as the linear coefficient of the mean variance of refractive index difference of two points over $r^{2/3}$, where $r$ denotes the spacing of the two points.     An open problem is that real measurement data for the RISC models in the THz band is still missing, and only theoretical RISC models are available~\cite{gao2024attenuationPIMRC}.    
    
    Next, after modeling the refractive index distribution in turbulence, we need to the turbulence-induced propagation loss. 
    For the THz single-input single-output (SISO) system equipped with directional antennas at both sides, the channel model can be simply extended from laser communications. However, these results can not be simply extended to the THz UM-MIMO system. 
    A recent analytical study on the attenuation model for THz UM-MIMO systems in turbulent medium demonstrates that the loss of spatial coherence in antenna arrays leads to additional propagation loss~\cite{gao2024attenuation}. 
    In summary, despite these theoretical analyses, the experimental results on the influence of turbulence on THz wave propagation are still missing.
        
    \subsection{Space Channel Modeling}
    The channel modeling in space communication scenarios is critical for inter-satellite communications. 
    At the boundary of the atmosphere and outer space, since medium molecules like water vapor and oxygen and air aerosols almost disappear, the space channels are unaffected by the molecular absorption effect, weather effect, and atmospheric turbulence. As a result, the plasma in the ionosphere generated by solar radiation becomes the dominant effect of THz wave propagation between satellites. Moreover, the influence of the extremely high velocity of satellites in space on the THz space channel needs investigation.
    
    Solar radiation permanently ionizes the particles of the outer surface of Earth, which forms the ionosphere of the atmosphere. These grouped charged particles, as a large ionospheric plasma, lead to plasma attenuation to THz wave propagation, When the charged particles in plasma are forced to oscillate upon the influence of the oscillating electric field of the EM wave, these particles absorb energy from the THz wave and results in attenuation.
    Based on the theory of forced oscillation, the relationship between the intrinsic oscillation frequency of the plasma (referred to as plasma frequency) and the frequency of the THz wave is the key factor determining the intensity of this oscillation, thus affecting the plasma attenuation.
    The plasma frequency is modeled in~\cite{nie2021channel}, which only depends on the electron density. In the THz band, the plasma attenuation monotonically decreases as the carrier frequency of THz signals increases, implying that the THz communications are more resistant to plasma attenuation than microwave and millimeter-wave communications. 
    \begin{figure}
        \centering
        \includegraphics[width=\linewidth]{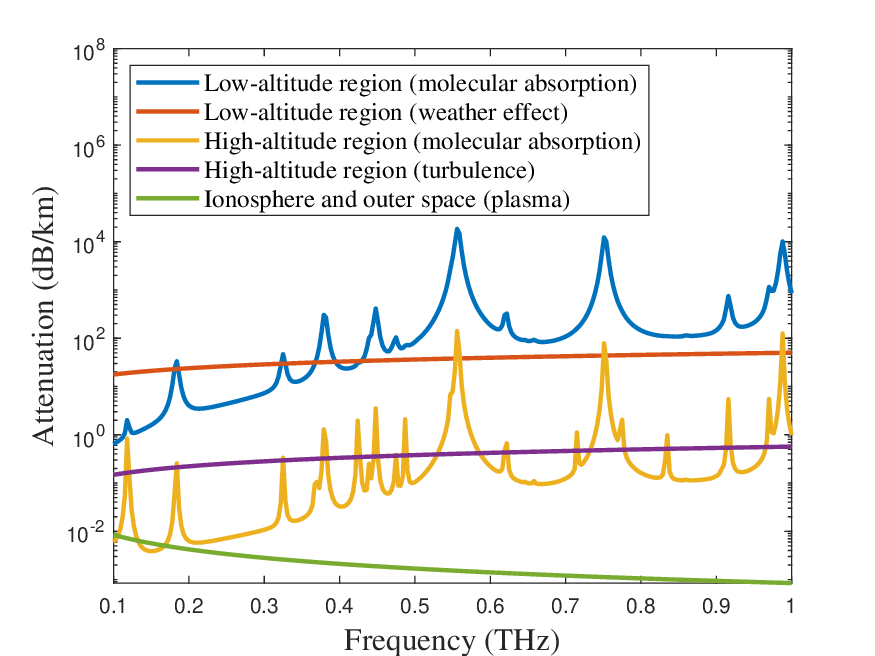}
        \caption{Attenuation effects in low-altitude regions, high-altitude regions, and outer space.}
        \label{fig:atten-comparison}
    \end{figure}
    Various attenuation effects in THz aerospace communications at various altitudes are compared in Fig.~\ref{fig:atten-comparison}. It is discovered that attenuation effects in lower-altitude regions are more severe than those in higher-altitude regions. Moreover, the frequency selectivity of the molecular absorption effect is more significant than that of other attenuation effects.

    Besides the attenuation caused by the ionosphere, the strong Doppler effect of inter-satellite space communications due to high velocity is another major challenge. Since the velocity of satellites may reach several tens of kilometers per second, the received signal experiences a severe Doppler effect. 
    Note that this Doppler effect in space communication is different from that in terrestrial communications. 
    In terrestrial communications, the received signal is combined from a lot of multi-paths, and different multi-path components from different angles of arrival experience distinct frequency shifts, resulting in a severe Doppler spread in the frequency domain.
    However, for LoS propagation in space communications, the received THz signal experiences a uniform Doppler effect, which means that a transmitted sine wave is still a sine wave at the receiver side, only with a constant frequency shift.
    Therefore, this Doppler frequency shift can be compensated by signal processing technique. The remaining challenge is to sense the velocity of satellites to aid this signal processing.

    \section{Enabling Technology for Terahertz Aerospace Communications}~\label{sec:enabling}
    After modeling the wireless channel, we describe some key enabling technologies for THz aerospace communications, including THz MIMO and intelligent surface, beam alignment and tracking, THz ICARS, and resource allocation. 
    \subsection{MIMO and Intelligent Surface}
    The high propagation loss of THz waves according to Friis' law motivates the use of highly directional antennas or large-scale antenna arrays. Since the wavelength is on the order of several millimeters, an ultra-massive antenna array still possesses an acceptable size, making THz UM-MIMO a promising and flexible technology to compensate for the spreading loss.
    From the perspective of system realization for UM-MIMO, the conventional fully-digital architecture of the UM-MIMO requires an equal number of RF chains as the antenna to flexibly control their amplitude and phase. However, this architecture is unrealistic since UAVs and satellites cannot carry such a large amount of communication devices. An alternative approach is to connect all antennas with one RF chain through phase shifters, but this method strongly limits the feasible beam pattern that can be steered as well as the achievable spectral efficiency. A hybrid beamforming architecture, which connects several RF chains with a large number of antennas, is instead widely used.  
    Recent advancements focus on developing a new architecture for antenna arrays and precoding algorithms for balanced hardware cost, power consumption, and spectral efficiency.

    Moreover, unlike terrestrial THz communications with many multi-paths from walls, humans, and vehicles in addition to the LoS path, space communications usually have only one LoS path. As a result, the multiplexing gain provided by THz UM-MIMO is limited by the spatial degrees of freedom. To this end, an intelligent surface, which manually creates additional multi-paths for aerospace communications, has been proposed to improve spectral efficiency.     
    Note that the near-field effect is not considered in aerospace communications because the propagation distance in aerospace usually exceeds the near-field boundary distance.
    
    \subsection{Beam Alignment and Tracking}
    THz pencil beams require high-precision alignment and tracking to fully provide the antenna gain, either generated by a directional antenna or UM-MIMO. 
    For directional antennas, such as a high-gain Cassegrain antenna, the beam direction is controlled by physically tuning the antenna orientation, but for THz UM-MIMO, the beam direction is electrically controlled by phase shifters. 
    For both approaches, a straightforward method of beam alignment and tracking is to constantly sense the wireless channel. However, due to the long distance between aerospace communication units, the propagation lag of communication signals behind the channel estimation leads to misalignment.     
    Therefore, given that trajectories of UAVs and satellites are pre-determined, modeling of their beam deviation patterns is critical to obtaining prior knowledge to assist with beam alignment and tracking. 

    The source of beam deviation is the small external forces on the satellites including the unrealistic gravitational errors and the solar radiation force.    
    Satellites move around a circular orbit under the gravity of the Earth. However, since the actual shape of Earth is approximately an ellipsoid, with an equatorial diameter larger than the polar diameter by about 44 km, the gravity exerted on the satellites is not ideally stable. This slight gravity difference results in small perturbations on the satellites, known as the $J_2$ effect. 
    Moreover, the solar radiation force is another factor contributing to the random perturbation of satellites. According to wave-particle duality, photons of the sunlight have a small amount of momentum, which exerts a small solar radiation force on the satellites when the photons are reflected or absorbed. 
    In~\cite{nie2021channel}, a deterministic analysis of the perturbation of satellites due to the $J_2$ effect and solar radiation is conducted, revealing that the beam angle deviation approximately follows a Gaussian distribution with zero mean and a standard deviation of $0.23~\textrm{rad}$.   
    
    \subsection{Integrated Communications and Radar Sensing (ICARS)}
\begin{figure*}
        \centering
        \includegraphics[width=0.8\linewidth]{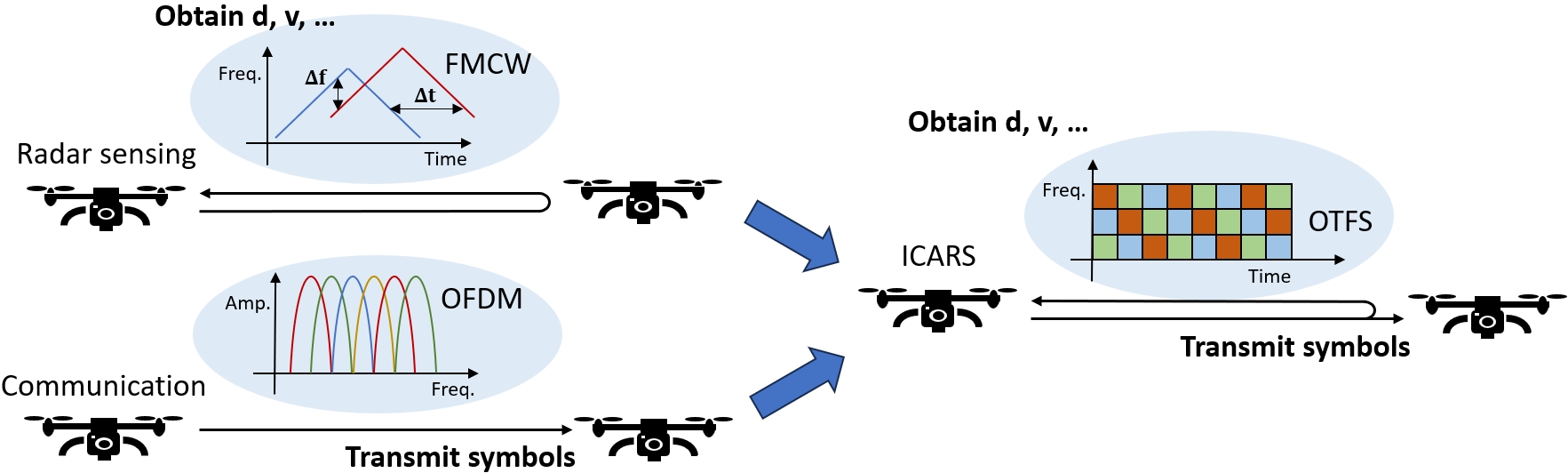}
        \caption{Illustration of OTFS signals as ICARS waveform.}
        \label{fig:OTFS}
    \end{figure*}

    THz sensing, due to the sub-millimeter-level THz wavelengths, has numerous potential applications in aerospace. 
    According to Planck's black-body radiation law, the peak radiation of interstellar gas and dust irradiated by thermal radiation at a temperature between $5~\textrm{K}$ and $100~\textrm{K}$ is in the THz band~\cite{siegel2010thz}. NASA has experimentally shown that nearly half of the cosmic radiation energy and 98\% of the photons emitted by the universe since the Big Bang are in the THz or far-infrared frequency band, highlighting the critical need for THz space detection. 
    Developing THz ICARS systems can reduce the hardware costs of joint systems and lessen the load on UAVs and satellites.
    In the THz ICARS module, the transmitter side sends a THz signal while the receiver side simultaneously performs detection and demodulation based on the received signal. 
    To achieve this, a joint waveform design and integrated signal processing tailored for these two functionalities are required, which are elaborated as follows:

    First, for waveform design, conventional OFDM waveforms have achieved significant success in 4G and 5G networks, but they have a high peak-to-average-power ratio (PAPR) and rely on orthogonality among sub-carriers in the frequency domain. Both of these features pose challenges for THz aerospace communication, which operates under limited transit power and high mobility.
    To address these challenges, researchers are exploring new waveforms based on OFDM, e.g., DFT-s-OFDM, OTFS, and DFT-s-OTFS~\cite{han2024thz}. Specifically, DFT-s-OFDM reduces the PAPR of OFDM by incorporating a DFT module into the OFDM system, while OTFS improves resilience against Doppler effects by transforming the modulated signal into the delay-Doppler domain, as shown in Fig.~\ref{fig:OTFS}. DFT-s-OTFS, which combines both improved versions of OFDM, offers low PAPR and enhanced resilience against Doppler effects. However, the hardware complexity of DFT-s-OTFS is on the high end, requiring additional computational resources.
    Given the complex environment of THz aerospace communications, these candidate waveforms should be carefully compared and selected for various applications. Further improvements to the ICARS waveform are still needed for THz aerospace communications.
    
    Second, for the design of receiver signal processing, the different tasks for communication and radar sensing result in significant differences in the signal processing approaches. The communication side aims to decode the modulated signal in the presence of thermal noise, while the sensing component seeks to extract information about the external environment based on the received signal. 
    Given the limited computational resources available on satellites or UAVs, there is a strong demand for a joint design of signal processing that reduces complexity. Recent advancements have focused on developing multi-task signal processing algorithms that simultaneously perform demodulation and detection tasks using artificial intelligence (AI) techniques. For example, a joint signal processing approach based on a multi-task neural network is proposed to carry out sensing, channel estimation, and equalization simultaneously, with a reduced computational complexity. 

    \subsection{Resource Allocation for THz Aerospace Network}
    Conventionally, on-board devices carried by satellites are designed for different specific applications, and these devices only handle their own computational tasks, which makes them unable to perform computation-demanding missions. 
    Nowadays, as the trends of transforming isolated satellites into cooperating satellite chains or constellations, computing-specialized satellites, such as edge-computing servers with powerful computing capabilities, are designed to manage the computing tasks for other client satellites with limited resources.
    On one hand, the allocation of computing tasks from client to server satellites, known as task offloading, requires centralizing management.
    On one hand, the task offloading process puts a higher throughput requirement on the communication network. 
    Additionally, as the number of satellites has increased dramatically in recent years, the constellation of satellite networks has become increasingly complex. Consequently, an efficient allocation mechanism for communication and computational resources is essential for THz aerospace networks. However, it has been found that the resource efficiency maximization problem for THz aerospace networks is NP-hard, indicating that a low-complexity and efficient closed-form solution is impractical using conventional optimization approaches. Deep reinforcement learning algorithms present a promising technology for addressing this complex planning problem. Extensive solutions are still needed.
    
    \section{Open Problems and Future Directions}~\label{sec:future}
    \subsection{Fusion of Multi-Frequency Systems}
    Despite the advantages of THz band communications over microwave and optical frequency bands in terms of spectral efficiency and tolerance to beam misalignment, these other frequency bands also offer distinct benefits. 
    For microwave communications, although the data rate is limited by small bandwidth, the wide beamwidth and low path loss contribute to greater robustness against deep fading.
    By integrating these lower-frequency bands with the THz band, microwave signals can improve the reliability of THz beam alignment and tracking.
    In the case of FSO communications, it has been found that THz signals and FSO signals exhibit different penetrating capabilities under various weather conditions. For example, THz signals experience negligible fog attenuation, low propagation loss in the presence of dust and turbulence, medium propagation loss during rain, and high molecular absorption loss due to water vapor in coastal regions. In contrast, FSO communications are susceptible to fog, dust, rain, and turbulence but are largely unaffected by water vapor. Therefore, a hybrid FSO/THz aerospace communication system can provide both high data rates and reliability across various communication scenarios.
   
    \subsection{Hardware Imperfectness in Space}
    In the last century, the THz band was considered a ``THz gap" due to the hardware limitations of THz transceivers.
    Recently, as THz device technologies have evolved, various methods for THz signal generation have been proposed. However, the remaining change is the imperfection of THz transceivers, such as quantization errors in analog-digital converters and digital-to-analog converters, IQ imbalance, phase noise, amplitude and phase errors in phases shifters, and non-linearity of power amplifiers and low-noise amplifiers. These hardware imperfections can be compensated by signal processing algorithms, e.g., some machine-learning algorithm~\cite{zhao2024dynamic}.  
    In addition, ultra-low temperature environments and cosmic rays from the universe may cause signal distortion on these devices.
    Since THz UM-MIMO requires precise phase control among a large number of antennas, the influence of these effects on THz devices needs to be extensively studied.

    \subsection{AI for THz Aerospace Communications}
    \begin{figure*}
        \centering
        \includegraphics[width=0.7\linewidth]{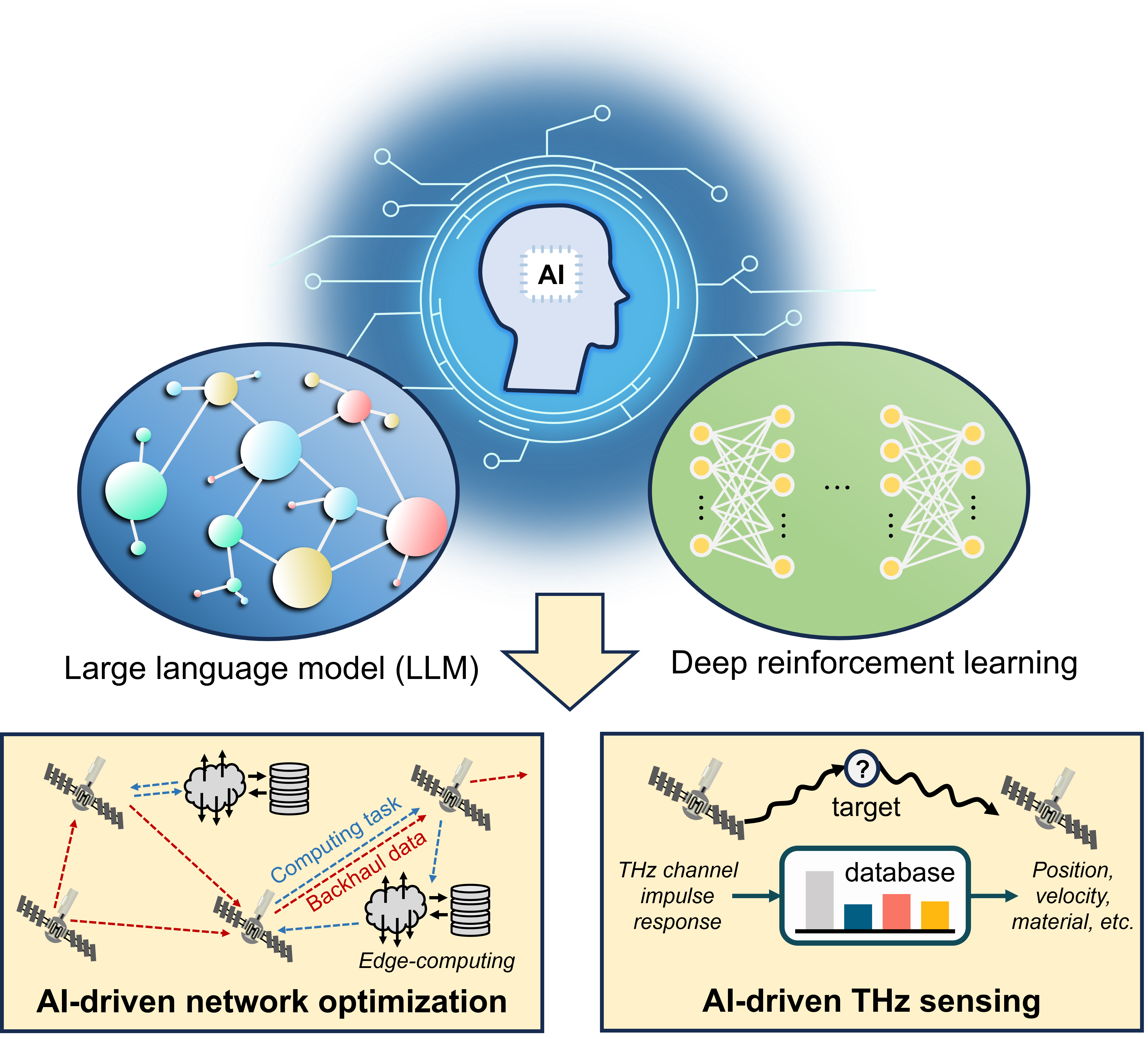}
        \caption{Future AI technologies for THz Aerospace communications.}
        \label{fig:ai}
    \end{figure*}
    AI technologies have been successfully utilized in diverse applications, including wireless communication~\cite{fourati2021artificial}. Two potential research directions are summarized in Fig.~\ref{fig:ai}.
    On one hand, large language model-aided data acquisition represents an important application of AI technology. AI-powered data acquisition aims to infer environmental and transceiver information from THz propagation characteristics. By leveraging a large language model trained on extensive datasets regarding the interaction between THz waves and the environment, various environmental data can be extracted from detected signals. 
    These AI-based data acquisition techniques hold potential for future sensing and detection applications, such as THz space debris detection and THz meteorological observations.
    
    On the other hand, to tackle the challenges posed by complex and rapidly changing satellite topologies, deep reinforcement learning techniques emerge as promising candidates for resource allocation. Specifically, the strong representation ability of deep neural networks, combined with reinforcement learning strategies, can effectively manage network topology complexity and adapt to frequent variations in satellite constellation.
    This approach addresses the intricate resource allocation issues in future aerospace networks, which consist of various functional devices, including edge-computing nodes, sensing satellites, and communication devices.
    \section{Conclusion}\label{sec:concl}
   In this paper, a comprehensive overview of THz aerospace communication is presented, covering channel modeling, key enabling technologies, and future research directions. Specifically, the THz aerospace channel for low-altitude regions, high-altitude regions, and outer space is described and analyzed. Then, the key enabling technologies including MIMO and intelligent surface, beam alignment and tracking, THz ICARS, and resource allocation for networking are elaborated. Finally, future research directions are summarized. 
	\bibliographystyle{IEEEtran}
	\bibliography{aerospace}
\end{document}